\documentclass[11pt,english]{article}
\usepackage{amsmath,amssymb,color,slashed,graphicx,babel}

\makeatletter
\@ifundefined{showcaptionsetup}{}{%
\PassOptionsToPackage{caption=false}{subfig}}
\usepackage{subfig}
\makeatother

\usepackage{amsfonts}

\newcommand{\hoch}[1]{$\, ^{#1}$}

\newcommand{\be}{\begin{equation}}
\newcommand{\ee}{\end{equation}}
\newcommand{\bea}{\setlength\arraycolsep{2pt} \begin{eqnarray}}
\newcommand{\eea}{\end{eqnarray}}

\def\ft#1#2{{\textstyle{\frac{\scriptstyle #1}{\scriptstyle #2} } }}
\def\fft#1#2{{\frac{#1}{#2}}}

\def\0{{\sst{(0)}}}
\def\1{{\sst{(1)}}}
\def\2{{\sst{(2)}}}
\def\3{{\sst{(3)}}}
\def\4{{\sst{(4)}}}
\def\5{{\sst{(5)}}}
\def\6{{\sst{(6)}}}
\def\7{{\sst{(7)}}}
\def\8{{\sst{(8)}}}
\def\sst#1{{\scriptscriptstyle #1}}

\begin{document}

\begin{flushright}
\hfill{CAS-KITPC/ITP-277}
\end{flushright}

\vspace{25pt}
\begin{center}
{\large {\bf Linearized Modes in Extended and Critical Gravities }}

\vspace{10pt}

Yi-Xin Chen\hoch{1}, H. L\"u\hoch{2,3} and Kai-Nan Shao\hoch{1,4}

\vspace{10pt}

\hoch{1}{\it Zheijiang Institute of Modern Physics\\
Department of Physics, Zheijiang University, Hangzhou 310027}

\vspace{10pt}

\hoch{2}{\it China Economics and Management Academy\\
Central University of Finance and Economics, Beijing 100081}

\vspace{10pt}

\hoch{3}{\it Institute for Advanced Study, Shenzhen University\\
Nanhai Ave 3688, Shenzhen 518060}

\vspace{10pt}

\hoch{4} {\it Kavli Institute for Theoretical Physics China, CAS,
Beijing 100190}

\vspace{40pt}

\underline{ABSTRACT}
\end{center}

We construct explicit solutions for the linearized massive and
massless spin-2, vector and scalar modes around the AdS spacetimes
in diverse dimensions. These modes may arise in extended
(super)gravities with higher curvature terms in general dimensions.
Log modes in critical gravities can also be straightforwardly
deduced.  We analyze the properties of these modes and obtain the
tachyon-free condition, which allows negative mass square for these
modes. However, such modes may not satisfy the standard AdS boundary
condition and can be truncated out from the spectrum.

\vspace{15pt}

\thispagestyle{empty}





\newpage

\section{Introduction}

One natural and simple method of resolving the renormalizability of
General Relativity is to extend the Einstein theory with
higher-order curvature terms; however, the procedure tends to
introduce ghost massive spin-2 modes, in addition to the usual
massless graviton \cite{stelle1,stelle2}.  The situation is improved
in $D=3$, since the massless graviton is pure gauge and hence the
theory can become unitary by reversing the sign of the
Einstein-Hilbert action. The examples include the celebrated
topologically massive gravity \cite{tmg} and the recently
constructed new massive gravity \cite{nmg}. When the cosmological
constant is included, there can exist a critical point in the
parameter space for which the unitarity can be achieved without
having to reverse the sign of the Einstein-Hilbert action
\cite{lss}.

    With recent interest and progress in three dimensions, it is
natural to extend the discussion to $D=4$.  Critical gravity in four
dimensions was proposed in \cite{lpcritical}. It improves
significantly the situation discussed in \cite{stelle1,stelle2} with
the introduction of a cosmological constant. The theory contains the
Einstein Hilbert action, the cosmological constant and a Weyl-square
term. It is however somewhat vacuous, in that many physical
quantities, such as the energy of the massless graviton, mass and
entropy of the Schwarzschild-AdS (Anti-de Sitter) black hole, vanish
identically at the critical point. These results are consistent with
the latest proposal \cite{maldaconf} that cosmological Einstein
gravity can emerge from conformal gravity, which involves only the
Weyl-square term, in the long-wavelength limit precisely at the
critical point. In a recent work \cite{lpp}, it was shown that the
relation between Einstein and conform gravities can be extended to
six dimensions, and hence likely in general even dimensions.
Furthermore, the unitarity in the Weyl-square extended gravity
\cite{lpcritical} can be achieved in a wider continuous parameter
region than just at the critical point \cite{lpp}. This provides a
much more likely candidate for renormalizable quantum gravity.  The
supersymmetric generalization of the critical gravity and its
unitary extension were obtained in \cite{lpsw}, analogous to the
off-shell extended supergravities in three dimensions
\cite{Andringa:2009yc}-\cite{Lu:2010cg}. Recent related works in
critical gravities can be found in \cite{liusun}-\cite{Moon:2011ef}

     Many important properties derived in extended gravities rely
on the explicit solutions of the massive and massless graviton
modes. For example, once we know the explicit form of a mode, we can
determine whether it is a ghost or not by evaluating its energy. We
can also determine whether the mode is consistent with the AdS
boundary condition.  It turns out that the tachyon-free condition,
{\it i.e.} the absence of exponential growth in time, allows
negative mass square in AdS spacetime even for spin-2 modes,
generalizing the Breitenlohner-Freedman bound for scalars. However,
these modes has a slower falloff at the AdS boundary. One can then
engineer the parameters such that the inevitable ghosts are these
massive modes, which can be truncated out by the standard boundary
condition, leaving only the massless graviton in the spectrum, which
can be evaluated to have positive energy \cite{lpp}. In this
analysis, the explicit solutions of the massive and massless
graviton modes play an important role. For future research in
extended gravities, it is therefore advantageous to construct the
linearized modes around the AdS spacetimes in general dimensions.
The task is made simpler by the fact that in many examples of
extended gravities, the differential operators acting on the modes
factorize and hence the general solutions are simply the linear
combinations of massless and massive modes with different masses.

   In this paper, we present linearized modes around the AdS vacua
that could arise in extended gravities in diverse dimensions.
(Examples of full non-linear pp-wave type of solutions in extended
gravities can be found in \cite{Alishahiha:2011yb,Gullu:2011sj}.) In
general, the spectrum contains the massive trace scalar, massless
and massive spin-2 modes.  The vector modes from the metric can be
removed by the diffeomorphism which is preserved in extended
gravities.  Around the AdS vacuum, the equation of motion for the
trace scalar mode, in a suitable gauge choice, involves two
derivatives less than the spin-2 modes \cite{lpcritical}, and hence
they are undesirable from the point of view of renormalizality.
(This conclusion differs from that in Minkowskian vacuum obtained in
\cite{stelle1,stelle2}.) Thus the trace mode in critical gravity is
typically decoupled from the system by a suitable parameter choice
\cite{lpcritical}. (See also \cite{nmg}.) If the trace mode were
present, it could provide a source for the the spin-2 modes in
higher-derivative theories, even at the linearized level. Although a
field redefinition of the spin-2 modes can remove the source
\cite{bhrst}, it involves derivatives and hence may not be
invertible. In this paper, we shall only consider the situation
where the spin-2 modes are decoupled from the trace mode. In this
case, the most general solutions are simply the linear combination
of massless and massive graviton modes (or log modes in the critical
points). Note that the (spin-0) scalar and vector (spin-1) matter
fields do however arise in Weyl-squared supergravity \cite{lpsw}.
Thus in this paper, we shall present the explicit construction of
the general linearized decoupled massive (massless) scalar, vector
and spin-2 graviton modes around the AdS vacua in diverse
dimensions.

        In section 2, we present the general formalism for the
construction.  The construction, which follows from \cite{lss,bhrt},
is based on the fact that the AdS$_D$ spacetime has $SO(2,D-1)$
isometry.  We can hence obtain the general modes by obtaining the
highest weight state of the $SO(2,D-1)$.  The full representation
can be obtained by the action of the creation generators of the
$SO(D-1)$ subgroup. In section 3, we obtain the explicit results for
the scalar, vector and spin-2 modes in $D=3$ and $D=4$.  We recover
the previously known spin-2 solutions in $D=3$ \cite{lss} and $D=4$
\cite{bhrt}. In section 4, we present the explicit new solutions for
AdS$_5$ and AdS$_6$.  The general solutions in an arbitrary
dimension are presented in section 5. In section 6, we give the
general formalism on how to obtain log modes at the critical point
from the solutions of general massive modes. The conclusion and
further discussions are given in section 7, where we give a
definition of the ``true mass'' of linearized modes in AdS
backgrounds.

\section{General formalism}

The $D$-dimensional anti-de Sitter spacetime (AdS$_D$) can be
represented as its embedding in the $(D+1)$-dimensional flat space
as a hyperboloid
\begin{equation}
x_1^2 + x_2^2 - \sum_{i=3}^D x_i^2=L^2\,,
\end{equation}
with the flat metric
\begin{equation}
ds^2 = -dx_1^2 - dx_2^2 + \sum_{i=3}^D dx_i^2\,.
\end{equation}
The global coordinates of the AdS spacetime is given by
\begin{eqnarray}
x_{1} = L\cosh\rho\,\cos\tau\,,&& x_{2}  =
L\cosh\rho\,\sin\tau\,,\cr 
x_{2a+1} = L\sinh\rho\, u_{a}\,\cos\phi_{a}\,,&& x_{2a+2}=
L\sinh\rho\, u_{a}\,\sin\phi_{a}\,,
\end{eqnarray}
where $L$ is the ``radius'' of the AdS, and the index $a$ runs from
1 to $[\fft{D}2]$, and
\begin{equation}
\sum_{a=1}^{[\fft{D}2]} u_a^2 = 1\,.
\end{equation}
Note that for odd D, the $\phi$ coordinate with the largest index
should be set to zero. The AdS$_D$ metric in the global coordinates
is given by
\begin{equation}
ds^{2}=L^{2}\Big[-\cosh^{2}\rho\,
d\tau^{2}+d\rho^{2}+\sinh^{2}\rho\Big(\sum_{a}^{[\fft{D}2]}
du_{a}^{2} +\sum_{a}^{[\fft{D-1}2]}u_{a}^{2}d\phi_{a}^{2}
\Big)\Big]\,, \label{eq:AdSnmetric}
\end{equation}
The metric is Einstein and maximally symmetric, namely
\begin{equation}
R_{\mu\nu}=\Lambda g_{\mu\nu}\,,\qquad
R_{\mu\nu\rho\sigma}=\fft{\Lambda}{D-1}(g_{\mu\rho} g_{\nu\sigma} -
g_{\mu \sigma} g_{\nu\rho})\,,
\end{equation}
with the cosmological constant $\Lambda =-(D-1)L^2$.

      The AdS$_D$ spacetime possesses the isometry group $SO(2,
D-1)$, generated by the following Killing vectors
\begin{eqnarray}
L_{12} = x_{1}\frac{\partial}{\partial x_{2}}-
x_{2}\frac{\partial}{\partial x_{1}}\,,&&
L_{ij}=x_{i}\frac{\partial}{\partial x_{j}}-
x_{j}\frac{\partial}{\partial x_{i}}\,,\cr 
L_{1i} = -{\rm i}\left(x_{1}\frac{\partial}{\partial x_{i}}+
x_{i}\frac{\partial}{\partial x_{1}}\right)\,,&& L_{2i} = -{\rm
i}\left(x_{2}\frac{\partial}{\partial
x_{i}}+x_{i}\frac{\partial}{\partial x_{2}}\right)\,.
\end{eqnarray}
These Killing vectors generate the following algebra
\begin{equation}
\left[L_{ij},L_{kl}\right]=-\delta_{ik}L_{jl}-\delta_{jl}L_{ik}
+\delta_{il}L_{jk}+\delta_{jk}L_{il}\,.\label{eq:Lijcommute}
\end{equation}
Note that in the above the indices include the 1, 2 directions as
well.

     Let us classify the Cartan and root generators.  When $D$ is
odd, namely $D=2n-1$.  The group $SO(2,2n-2)$ belongs to Lie group
$D_n$. The root vectors for the simple roots are given by
\begin{eqnarray}
\vec{\alpha}_{i} & = & \vec{e}_{i}-\vec{e}_{i+1}\qquad1\leq i\leq
n-1\,,\cr
\vec{\alpha}_{n} & = & \vec{e}_{n-1}+\vec{e}_{n}\,,
\end{eqnarray}
where $\vec{e}_{1}=\left(-1,0,\cdots,0\right)$ and
$\vec{e}_{i}=\left(0,\cdots,1,\cdots,0\right)$ for $i\ne 1$.  Note
that the unusual sign choice for $\vec e_1$, which is convenient for
the non-compact group.  A natural choice of the Cartan generators
for our AdS metrics in global coordinates is given by
\begin{equation}
H_{i}={\rm i}\, L_{(2i-1)(2i)}\,,\qquad i=1,2,\ldots, n\,.
\end{equation}
The corresponding simple roots are then given by
\begin{eqnarray}
E_{\vec{\alpha}_{1}} & = & \ft12\left(L_{13}+{\rm i}\,L_{14}+{\rm
i}\,L_{23}-L_{24}\right)\,,\cr E_{\vec{\alpha}_{i}} & = &
\ft12\left(L_{(2i-1)(2i+1)} +{\rm i}\,L_{(2i-1)(2i+2)}-{\rm i}\,
L_{(2i)(2i+1)}+L_{(2i)(2i+2)}\right)\,, \quad 2\le i\le n-1\,,\cr 
E_{\vec{\alpha}_{n}} & = & \ft12\left(L_{(2n-3)(2n-1)}-{\rm
i}\,L_{(2n-3)(2n)}- {\rm
i}\,L_{(2n-2)(2n-1)}-L_{(2n-2)(2n)}\right)\,.
\end{eqnarray}
It is easy to verify that they satisfy
\begin{eqnarray}
\left[H_{i},H_{j}\right] & = & 0,\qquad
\left[H_{i},E_{\vec{\alpha}_{x}}\right]
=\vec{\alpha}_{x}^{i}E_{\vec{\alpha}_{j}}\,,\qquad
x=1,\dots,n\label{eq:SOnalgebra}\\
\left[E_{\vec{\alpha}_{x}},E_{-\vec{\alpha}_{x}}\right] & = &
\frac{2}{\left|\vec{\alpha}_{x}\right|^{2}}\vec{\alpha}_{x}^{i}H_{i}
\,,\label{eq:SOnalgebra2}
\end{eqnarray}
where
\begin{equation}
\left|\vec{\alpha}_{x}\right|^{2}=\sum_{i=1}^{n}
\left(\vec{\alpha}_{x}^{i}\right)^{2}\,.
\end{equation}
All the negative (simple) roots are then given by
$E_{-\vec{\alpha}_{x}}=\pm\left(E_{\vec{\alpha}_{x}}\right)^{*}$.
(Note that the specific choice of the $\pm$ signs in the negative
roots are determined according to the algebraic relation
(\ref{eq:SOnalgebra2}).)

    For $D=2n$, the isometry group $SO(2,2n-1)$ belongs to the $B_n$
series. The root vectors for the simple roots are
\begin{eqnarray}
\vec{\alpha}_{i} & = & \vec{e}_{i}-\vec{e}_{i+1}\,,\quad1\leq i\leq
n-1\,,\cr 
\vec{\alpha}_{n} & = & \vec{e}_{n}\,.
\end{eqnarray}
The Cartan generators and the positive roots are chosen to be
\begin{equation}
H_{i}={\rm i}\,L_{(2i-1)(2i)}\,,\qquad i=1,2,\ldots,n\,,
\end{equation}
and
\begin{eqnarray}
E_{\vec{\alpha}_{1}} & = & \ft12\left(L_{13}+{\rm i}\,L_{14}+ {\rm
i}\,L_{23}-L_{24}\right)\,,\cr 
E_{\vec{\alpha}_{i}} & = & \ft12\left(L_{(2i-1)(2i+1)}+{\rm
i}\,L_{(2i-1)(2i+2)}- {\rm
i}\,L_{(2i)(2i+1)}+L_{(2i)(2i+2)}\right)\,, \quad 2\le i\le
n-1\,,\cr 
E_{\vec{\alpha}_{n}} & = & L_{(2n-1)(2n+1)}-{\rm
i}\,L_{(2n)(2n+1)}\,.
\end{eqnarray}
They also satisfy the commutation relations (\ref{eq:SOnalgebra}),
(\ref{eq:SOnalgebra2}).

     In this paper, we are looking for solutions that are
eigenstates of Cartan generators and annihilated by all the
positive-root generators, namely
\begin{eqnarray}
H_1 |\psi\rangle &=& E_0 |\psi\rangle\,,\qquad H_i |\psi\rangle =
s_{i-1} |\psi\rangle\qquad \hbox{for}\qquad i=2,3,\ldots\,,\cr 
E_{\vec \alpha_x} |\psi\rangle &=& 0\,, \qquad \hbox{for all simple
roots $\alpha_x$ and hence all positive roots.}\label{geneom}
\end{eqnarray}
Note that the generators act on the solution as Lie derivatives.
Here $\psi$ includes the scalar $\Phi$, vector $A_\mu$ and spin-2
modes $h_{\mu\nu}$, satisfying
\begin{equation}
(\Box -M_0^2)\Phi = 0\,,\quad (\Box + \fft{1}{L^2} -M_1^2)A_\mu =
0\,,\quad (\Box +\fft{2}{L^2} - M_2^2) h_{\mu\nu} =
0\,.\label{spin012eom}
\end{equation}
The vector $A_\mu$ satisfies the transverse condition
\begin{equation}
\nabla^\mu A_\mu=0\,,\label{at}
\end{equation}
and $h_{\mu\nu}$ is both traceless and transverse
\begin{equation}
g^{\mu\nu}h_{\mu\nu} = 0\,,\qquad \nabla^\mu
h_{\mu\nu}=0\,.\label{htt}
\end{equation}
The reason we consider the vector modes in this paper is that
massive vectors naturally appear in critical supergravities in the
Proca form, satisfying $d{*F} =c\,  {*A}$. The transverse condition
(\ref{at}) follows straightforwardly. (See \cite{lpsw}.)

The equations in (\ref{spin012eom}) can be summarized in the unified
form for all spin-$s$ fields
\begin{equation}
(\Box + \fft{s}{L^2} - M_s^2) \psi_s =0\,.\label{spinseom}
\end{equation}
Not that in this formalism, the Casimir operator ${\cal E}$ is
related to the covariant Laplacian operator $\Delta$ as follows
\begin{equation}
{\cal E} = \sum_i H_i H_i + \sum_{x} \ft12 |\alpha_x|^2
(E^{\alpha_x} E^{-\alpha_x} +E^{-\alpha_x} E^{\alpha_x}) = -
L^2\Delta\,,
\end{equation}
where two sums are over all the Cartan and root generators
respectively.  Thus the solutions of (\ref{geneom}) must also
satisfy the equations in (\ref{spin012eom}), with appropriate $M_0$,
$M_1$ and $M_2$.

   Having obtained the highest weight state, we can obtain the
remaining modes by acting on the state with the negative root
generators in the subgroup $SO(D-1)$ of the $SO(2,D-1)$.  Thus the
massive modes with the same $E_0$ and hence the same $M_s^2$ form
scalar, vector and spin-2 representations of $SO(D-1)$,
corresponding to have 1, $D-1$ and $\ft12 D(D-1)-1$ degrees of
freedom respectively. In the massless limit, the scalar, vector and
spin-2 modes form representations of $SO(D-2)$, corresponding to
$D-2$ and $\ft12(D-1)(D-2)-1$ degrees of freedom, owing to the
additional gauge symmetry.  Note that this procedure for getting the
linearized spin-2 modes were spelled out in \cite{lss} for
three-dimensions and in \cite{bhrt} for four-dimensions.

      It should be pointed out that $M_s^2$ defined in
(\ref{spinseom}) is only the mass parameter.  To determine the true
mass of a mode, it is necessary to determine first the proper
definition of the masslessness.  For the spin-2 graviton, it follows
from the Einstein theory that the massless graviton corresponds
indeed to $M_2=0$. The massless vector $(s=1)$ can be defined that
it preserves the gauge symmetry, and hence $d{*dA}=0$, corresponding
to
\begin{equation}
(\Box + \fft{D-1}{L^2})A_\mu^{(0)}=0\,.
\end{equation}
The masslessness of a scalar field in AdS is more subtle, since
there is no difference from the point of view of symmetry between
the massive and massless scalars.  However, as we shall see in
section 7, by studying the falloff behavior, a massless scalar can
be nevertheless defined.

In order to find the solutions in global coordinates, we further
transform the generators $H_i$ and $E_{\vec \alpha_x}$ into the
expressions in terms of the global coordinates of the AdS metric
(\ref{eq:AdSnmetric}). We shall give the explicit expressions of
these generators in the low-lying dimensions in the follow sections.

\section{Linearized modes in AdS$_3$ and AdS$_4$}

\subsection{AdS$_3$}

The AdS$_{3}$ metric is given by setting $u_{1}=1$ in the general
AdS metric (\ref{eq:AdSnmetric}), namely
\begin{equation}
ds^{2}=L^{2}\left(-\cosh^{2}\rho\,
d\tau^{2}+d\rho^{2}+\sinh^{2}\rho\, d\phi^{2}\right)\,.
\end{equation}
The isometry group of the AdS$_{3}$ is $SO(2,2)$. There are two
Cartan generators
\begin{equation}
H_{1}={\rm i} L_{12}={\rm i}\frac{\partial}{\partial\tau}\,,\qquad
H_{2}={\rm i} L_{34}={\rm i} \frac{\partial}{\partial\phi}\,.
\end{equation}
The simple roots are $\vec{\alpha}_{1}=\left(-1,-1\right)$,
$\vec{\alpha}_{2}=\left(-1,1\right)$, and the corresponding
generators are
\begin{eqnarray}
E_{\vec{\alpha}_{1}} & = & \ft{1}{2}\left(L_{13}+{\rm i} L_{14}+{\rm
i} L_{23}-L_{24}\right)\cr &=& \ft{1}{2}e^{i\left(\tau+\phi\right)}
\tanh\rho\frac{\partial}{\partial\tau}-\ft{1}{2} {\rm i} e^{{\rm
i}\left(\tau+\phi\right)}\frac{\partial}{\partial\rho}+
\ft{1}{2}e^{{\rm i} \left(\tau+\phi\right)}\coth\rho
\frac{\partial}{\partial\phi}\,,\cr 
E_{\vec{\alpha_{2}}} &=&
\ft{1}{2}\left(L_{13}-{\rm i}L_{14}+{\rm i} L_{23}+L_{24}\right)\cr
&=& \ft{1}{2}e^{{\rm i}\left(\tau-\phi\right)}\tanh\rho
\frac{\partial}{\partial\tau}-\ft{1}{2}{\rm i} e^{{\rm
i}\left(\tau-\phi\right)}\frac{\partial}{\partial\rho}-\ft{1}{2}
e^{{\rm i}\left(\tau-\phi\right)}\coth\rho
\frac{\partial}{\partial\phi}\,.
\end{eqnarray}
The corresponding negative roots are given by
\begin{equation}
E_{-\vec{\alpha}_{1}}=\left(E_{\vec{\alpha}_{1}}\right)^{*}\,,\qquad
E_{-\vec{\alpha}_{2}}=\left(E_{\vec{\alpha}_{2}}\right)^{*}\,.
\end{equation}

\bigskip
\noindent{\bf Scalar modes:}
\medskip

We look for a scalar mode $\Phi(\tau,\rho,\phi)$ that forms the
highest weight state of the $SO(2,2)$ algebra. This is an eigenstate
of the Cartan generators $H_{1}$ and $H_{2}$, with eigenvalues $E_{0}$
and $s$, {\it i.e.}
\begin{equation}
H_{1}\Phi=E_{0}\Phi\,,\qquad H_{2}\Phi=s\Phi\,,
\end{equation}
while it is annihilated by the positives root generators $E_{\vec{\alpha}_{x}}$
\begin{equation}
E_{\vec{\alpha}_{x}}\Phi=0\,,\qquad x=1,2\,.
\end{equation}
We find that a nontrivial solution arises only when $s=0$; it is
given by
\begin{equation}
\Phi=e^{-{\rm i} E_{0}\tau}\left(\cosh\rho\right)^{-E_{0}}\,,
\end{equation}
It is easy to verify that the Laplacian action on $\Phi$
\begin{equation}
\Delta\Phi=-\Box\Phi=-\nabla_{\mu}\nabla^{\mu}\Phi=
-\frac{E_{0}\left(E_{0}-2\right)}{L^{2}}\Phi\,.
\end{equation}

\bigskip
\noindent{\bf Vector modes:}
\medskip

    The vector modes $A_{\mu}(\tau,\rho,\phi)$ that we are looking for
also belong to the highest weight state of $SO(2,2)$ algebra,
defined by
\begin{eqnarray}
H_{1}A_{\mu} & = & E_{0}A_{\mu}\,,\qquad
H_{2}A_{\mu}=sA_{\mu}\label{eq:AdS3_vector_H1H2}\,,\cr
E_{\vec{\alpha}_{x}}A_{\mu} & = & 0\,,\qquad
x=1,2\,.\label{eq:AdS3_vector_Eai}
\end{eqnarray}
We find that non-trivial solution may arise when $s=\pm 1$, given by
\begin{eqnarray}
A_{\tau} & = & e^{-{\rm i}\left(E_{0}\tau\pm\phi\right)}
\left(\cosh\rho\right)^{-E_{0}}\sinh\rho\,,\cr A_{\rho} & = &
i\left(\cosh\rho\right)^{-1}\left(\sinh\rho\right)^{-1}A_{\tau}\,,\cr
A_{\phi} & = & \pm A_{\tau}\,,
\end{eqnarray}
The gauge condition (\ref{at}) is automatically satisfied. The box
and Laplacian acting on the solutions are given by
\begin{eqnarray}
\Box A_{\mu}&=&\frac{E_{0}^{2}-2E_{0}-1}{L^{2}} A_{\mu}\,,
\label{eq:AdS3_vector_Box}\\
\Delta A_{\mu}&=&-\Box A_{\mu}+R_{\mu\nu}A^{\nu} =
-\frac{\left(E_{0}-1\right)^{2}}{L^{2}}A_{\mu}\,.
\label{eq:AdS3_vector_Laplacian}
\end{eqnarray}

\bigskip
\noindent{\bf Spin-2 modes:}
\medskip

The spin-2 modes $\psi_{\mu\nu}(\tau,\rho,\phi)$ that belong to the
highest weight state of $SO(2,2)$ algebra are defined by
\begin{eqnarray}
H_{1}\psi_{\mu\nu} & = & E_{0}\psi_{\mu\nu}\,, \qquad
H_{2}\psi_{\mu\nu}=s\psi_{\mu\nu}\,,\cr
E_{\vec{\alpha}_{x}}\psi_{\mu\nu} & = & 0\,,\qquad
x=1,2\,.\label{eq:AdS3_graviton_Eai}
\end{eqnarray}
The modes also satisfy the traceless and transverse conditions
\begin{eqnarray}
g^{\mu\nu}\psi_{\mu\nu}=0\,,\qquad \nabla^{\mu}\psi_{\mu\nu}=
0\,.\label{eq:AdS3tracelesstransverse}
\end{eqnarray}

     Non-trivial solutions arise when $s=\pm 2$, given by
\begin{eqnarray}
\psi_{\tau\tau} & = & e^{-{\rm i}\left(E_{0}\tau\pm
2\phi\right)}\left(\cosh\rho\right)^{-E_{0}}
\left(\sinh\rho\right)^{2}\,,\cr \psi_{\tau\rho} & = & {\rm
i}\left(\cosh\rho\right)^{-1}\left(\sinh\rho\right)^{-1}
\psi_{\tau\tau}\,,\cr \psi_{\tau\phi} & = & \pm
\psi_{\tau\tau}\,,\qquad \psi_{\rho\rho} = {\rm
i}\left(\cosh\rho\right)^{-1}
\left(\sinh\rho\right)^{-1}\psi_{\tau\rho}\,,\cr \psi_{\rho\phi} & =
& \pm \psi_{\tau\rho}\,,\qquad \psi_{\phi\phi} = \psi_{\tau\tau}\,.
\end{eqnarray}
The traceless and transverse conditions
(\ref{eq:AdS3tracelesstransverse}) are automatically satisfied. The
box and Laplacian operators acting on the solutions are given by
\begin{eqnarray}
\Box\psi_{\mu\nu}&=&\frac{E_{0}^{2}-2E_{0}-2}{L^{2}}\psi_{\mu\nu}\,,\cr
\Delta\psi_{\mu\nu} & = &
-\Box\psi_{\mu\nu}-2R^{\rho}{}_{\mu\sigma\nu}\psi_{\rho}{}^{\sigma}
+2R_{(\mu}{}^{\rho}\psi_{\nu)\rho}\cr & = &
-\frac{E_{0}^{2}-2E_{0}+4}{L^{2}}\psi_{\mu\nu}\,.
\end{eqnarray}
These spin-2 modes are precisely the ones given in \cite{lss}, with
$E_0=h + \bar h$ and $s=h-\bar h$, where the $SO(2,2)$ is
interpreted as $SL(2,R)\times SL(2,R)$.

    Note that we have followed the convention of \cite{lss} that the
complex solution for the spin-2 modes is denoted as $\psi_{\mu\nu}$,
with the understanding that the $h_{\mu\nu}$, which must be real, is
the real or imaginary part of $\psi_{\mu\nu}$.  However, we use the
same notation $\Phi$ and $A_\mu$ to denote the corresponding complex
solutions to economize the notations.

\subsection{AdS$_4$}

Letting $u_{1}=\sin\theta$, $u_{2}=\cos\theta$ in the general metric
(\ref{eq:AdSnmetric}), we have the standard AdS$_4$ metric in global
coordinates
\begin{equation}
ds^{2}=L^{2}\left(-\cosh^{2}\rho\,
d\tau^{2}+d\rho^{2}+\sinh^{2}\rho\left(d\theta^{2}+\sin^{2}\theta\,
d\phi^{2}\right)\right)\,.
\end{equation}
The isometry group is $SO(2,3)$. There are two Cartan generators
\begin{equation}
H_{1}={\rm i} L_{12}={\rm i} \frac{\partial}{\partial\tau}\,, \qquad
H_{2}={\rm i} L_{34}={\rm i} \frac{\partial}{\partial\phi}\,.
\end{equation}
The two simple roots are $\vec{\alpha}_{1}=\left(-1,-1\right)$,
$\vec{\alpha}_{2}=\left(0,1\right)$, and the remaining two positive
roots $\vec{\alpha}_{3}=\left(-1,0\right)$,
$\vec{\alpha}_{4}=\left(-1,1\right)$. The corresponding generators
are
\begin{eqnarray}
E_{\vec{\alpha}_{1}} & = & \ft{1}{2}\left(L_{13}+{\rm i} L_{14}+{\rm
i}L_{23}-L_{24}\right)\cr & = & \ft{1}{2}e^{{\rm
i}\left(\tau+\phi\right)}\sin\theta\tanh\rho
\frac{\partial}{\partial\tau}-\ft12 {\rm i} e^{{\rm i}
\left(\tau+\phi\right)} \sin\theta\frac{\partial}{\partial\rho}\cr
&& -\ft{1}{2}{\rm i}e^{{\rm
i}\left(\tau+\phi\right)}\cos\theta\coth\rho
\frac{\partial}{\partial\theta}+\ft{1}{2}e^{{\rm i}
\left(\tau+\phi\right)}
\csc\theta\coth\rho\frac{\partial}{\partial\phi}\,,\cr
E_{\vec{\alpha_{2}}} & = & L_{35}-{\rm i}L_{45}\cr
 & = & -e^{-i\phi}\frac{\partial}{\partial\theta}+{\rm i} e^{-{\rm i}\phi}
 \cot\theta\frac{\partial}{\partial\phi}\,,\cr
E_{\vec{\alpha}_{3}} & = & L_{15}+{\rm i} L_{25}\cr
 & = & e^{{\rm i}\tau}\cos\theta\tanh\rho\frac{\partial}{\partial\tau}
 -{\rm i}e^{{\rm i}\tau}\cos\theta\frac{\partial}{\partial\rho}+
{\rm i}e^{{\rm i}\tau}\sin\theta\coth\rho
\frac{\partial}{\partial\theta}\,,\cr E_{\vec{\alpha}_{4}} & = &
\ft{1}{2}\left(L_{13}-{\rm i}L_{14}+{\rm i}L_{23}+L_{24}\right)\cr
 & = & \ft{1}{2}e^{{\rm i}\left(\tau-\phi\right)}\sin\theta\tanh\rho
 \frac{\partial}{\partial\tau}-\ft{1}{2}{\rm i}e^{{\rm i}
\left(\tau-\phi\right)} \sin\theta\frac{\partial}{\partial\rho}\cr
&& -\ft{1}{2}{\rm i}e^{{\rm i} \left(\tau-\phi\right)}
\cos\theta\coth\rho \frac{\partial}{\partial\theta}-
\ft{1}{2}e^{{\rm i}\left(\tau-\phi\right)}
\csc\theta\coth\rho\frac{\partial}{\partial\phi}\,.
\end{eqnarray}
Note that $\left[E_{\vec{\alpha}_{1}},E_{\vec{\alpha}_{2}}\right]
=E_{\vec{\alpha}_{3}}$, $\left[E_{\vec{\alpha}_{2}},
E_{\vec{\alpha}_{3}}\right] =2E_{\vec{\alpha}_{4}}$. All the
negative-root generators are given by
\begin{equation}
E_{-\vec{\alpha}_{1}}=\left(E_{\vec{\alpha}_{1}}\right)^{*}\,,\quad
E_{-\vec{\alpha}_{2}}=-\left(E_{\vec{\alpha}_{2}}\right)^{*}\,,\quad
E_{-\vec{\alpha}_{3}}=\left(E_{\vec{\alpha}_{3}}\right)^{*}\,,\quad
E_{-\vec{\alpha}_{4}}=\left(E_{\vec{\alpha}_{4}}\right)^{*}\,.
\end{equation}
The generators for for the $SO(3)$ subgroup of the $SO(2,3)$ are
$H_2, E_{\vec \alpha_2}$ and $E_{-\vec \alpha_2}$.

\bigskip
\noindent{\bf Scalar modes:}
\medskip

     The scalar mode associated with the highest weight state of
$SO(2,3)$ has $s=0$ and it is given by
\begin{equation}
\Phi=e^{-{\rm i} E_{0}\tau}\left(\cosh\rho\right)^{-E_{0}}\,.
\end{equation}
The Laplacian action on $\Phi$ is
\begin{equation}
\Delta\Phi=-\Box\Phi=-\nabla_{\mu}\nabla^{\mu}\Phi=
-\frac{E_{0}\left(E_{0}-3\right)}{L^{2}}\Phi\,.
\end{equation}
It is clear that the scalar is a singlet under the $SO(3)$ subgroup.

\bigskip
\noindent{\bf Vector modes:}
\medskip

      The non-trivial solution has $s=1$, given by
\begin{eqnarray}
A_{\tau} & = & e^{-{\rm i}\left(E_{0}\tau+\phi\right)}\sin\theta
\left(\cosh\rho\right)^{-E_{0}}\sinh\rho\,,\cr A_{\rho} & = & {\rm
i}\left(\cosh\rho\right)^{-1}\left(\sinh\rho\right)^{-1}A_{\tau}\,,\cr
A_{\theta} & = & {\rm i}\left(\sin\theta\right)^{-1}
\left(\cos\theta\right)A_{\tau}\,,\cr A_{\phi} & = & A_{\tau}\,.
\end{eqnarray}
The gauge condition (\ref{at}) is automatically satisfied. The
action of the box and Laplacian operators on the solution is given
by
\begin{equation}
\Box A_{\mu}=\frac{E_{0}^{2}-3E_{0}-1}{L^{2}}A_{\mu}\,,\qquad
\Delta A_{\mu}=-\frac{\left(E_{0}-1\right)
\left(E_{0}-2\right)}{L^{2}}A_{\mu}\,.
\end{equation}
Acting repeatedly on the highest-weight state ($s=1$) with the
creation generator $E_{-\vec \alpha_2}$, we obtain the $s=0,-1$
modes as well.  Together, they form the spin-1 representation of
$SO(3)$.

\bigskip
\noindent{\bf Spin-2 modes:}
\medskip

In this case, we must have $s=2$ and the solution is given by
\begin{eqnarray}
\psi_{\tau\tau} &=& e^{-{\rm
i}\left(E_{0}\tau+2\phi\right)}\left(\sin\theta\right)^{2}
\left(\cosh\rho\right)^{-E_{0}}\left(\sinh\rho\right)^{2}\,,\cr
\psi_{\tau\rho} & = & {\rm i}\left(\cosh\rho\right)^{-1}
\left(\sinh\rho\right)^{-1}\psi_{\tau\tau}\,,\qquad
\psi_{\tau\theta} = i\left(\sin\theta\right)^{-1}
\left(\cos\theta\right)\psi_{\tau\tau}\,,\cr \psi_{\tau\phi} &=&
\psi_{\tau\tau}\,,\qquad \psi_{\rho\rho} = {\rm
i}\left(\cosh\rho\right)^{-1}
\left(\sinh\rho\right)^{-1}\psi_{\tau\rho}\,,\cr \psi_{\rho\theta}
&=& {\rm i}\left(\sin\theta\right)^{-1}
\left(\cos\theta\right)\psi_{\tau\rho}\,,\qquad \psi_{\rho\phi} =
\psi_{\tau\rho}\,,\cr \psi_{\theta\theta} & = & {\rm
i}\left(\sin\theta\right)^{-1}
\left(\cos\theta\right)\psi_{\tau\theta}\,,\qquad \psi_{\theta\phi}
= \psi_{\tau\theta}\,,\qquad \psi_{\phi\phi}= \psi_{\tau\tau}\,.
 \end{eqnarray}
The box and Laplacian action on this solution is given by
\begin{equation}
\Box\psi_{\mu\nu}=\frac{E_{0}^{2}-3E_{0}-2}{L^{2}}\psi_{\mu\nu}\,,\qquad
\Delta\psi_{\mu\nu}= -\frac{E_{0}^{2}-3E_{0}+6}{L^{2}}\psi_{\mu\nu}\,.
\end{equation}
The spin-2 massive modes were previously obtained in \cite{bhrt}. As
pointed out in \cite{bhrt}, acting on this highest state with
$E_{-\vec\alpha_2}$, one obtains the $s=1,0,-1,-2$ modes as well.
Together they form the spin-2 representation of $SO(3)$.

\section{Linearized modes in AdS$_5$ and AdS$_6$}

Although in the next section, we present linearized modes in AdS
spacetimes in general dimensions, we nevertheless give explicit
results for AdS$_5$ and AdS$_6$ in this section, since these
low-lying examples are more applicable than higher-dimensional ones.
In both cases, there are three Cartan generators, and hence we group
these two examples together.

\subsection{AdS$_5$}

The AdS$_{5}$ metric is given by setting $u_{1}=\sin\theta$,
$u_{2}=\cos\theta$ in the general AdS metric (\ref{eq:AdSnmetric}),
namely
\begin{equation}
ds^{2}=L^{2}\left(-\cosh^{2}\rho
d\tau^{2}+d\rho^{2}+\sinh^{2}\rho\left(d\theta^{2}+\sin^{2}\theta
d\phi_{1}^{2}+\cos^{2}\theta d\phi_{2}^{2}\right)\right)\,.
\end{equation}
The isometry group is $SO(2,4)$. The three Cartan generators
are
\begin{equation}
H_{1}={\rm i} L_{12}={\rm i}\frac{\partial}{\partial\tau}\,,\qquad
H_{2}={\rm i}L_{34}={\rm i}
\frac{\partial}{\partial\phi_{1}}\,,\qquad H_{3}={\rm i}L_{56}={\rm
i}\frac{\partial}{\partial\phi_{2}}\,.
\end{equation}
The simple roots are $\vec{\alpha}_{1}=\left(-1,-1,0\right)$,
$\vec{\alpha}_{2}=\left(0,1,-1\right)$,
 $\vec{\alpha}_{3}=\left(0,1,1\right)$. The corresponding generators
are
\begin{eqnarray}
E_{\vec{\alpha}_{1}} & = & \ft{1}{2}\left(L_{13}+iL_{14}+{\rm i}
L_{23}-L_{24}\right)\cr & = & \ft{1}{2}e^{{\rm
i}\left(\tau+\phi_{1}\right)}\sin\theta\tanh\rho
\frac{\partial}{\partial\tau}-\ft{1}{2}{\rm i}e^{{\rm i}\left(\tau+
\phi_{1}\right)}\sin\theta\frac{\partial}{\partial\rho}\cr &&
-\ft{1}{2}{\rm i}e^{i\left(\tau+\phi_{1}\right)}\cos\theta\coth\rho
\frac{\partial}{\partial\theta}+\ft{1}{2} e^{{\rm
i}\left(\tau+\phi_{1}\right)}\csc\theta\coth\rho
\frac{\partial}{\partial\phi_{1}}\,,\cr E_{\vec{\alpha_{2}}} &=&
\ft{1}{2}\left(L_{35}+{\rm i}L_{36}-{\rm i}L_{45}+L_{46}\right)\cr
&=&-\ft{1}{2}e^{-{\rm i}\left(\phi_{1}-\phi_{2}\right)}
\frac{\partial}{\partial\theta}+\ft{1}{2}{\rm i} e^{-{\rm
i}\left(\phi_{1}-\phi_{2}
\right)}\cot\theta\frac{\partial}{\partial\phi_{1}}+ \ft{1}{2}{\rm
i}e^{{\rm i}\left(\phi_{1}-\phi_{2}\right)}\tan\theta
\frac{\partial}{\partial\phi_{2}}\,,\cr E_{\vec{\alpha}_{3}} & = &
\ft{1}{2}\left(L_{35}-{\rm i}L_{36}-{\rm i}L_{45}-L_{46}\right)\cr &
= & -\ft{1}{2}e^{-{\rm i}\left(\phi_{1}+\phi_{2}\right)}
\frac{\partial}{\partial\theta}+\ft{1}{2}{\rm i}e^{-{\rm
i}\left(\phi_{1}
+\phi_{2}\right)}\cot\theta\frac{\partial}{\partial\phi_{1}}
-\ft{1}{2}{\rm i} e^{-{\rm i}\left(\phi_{1}+\phi_{2}\right)}
\tan\theta\frac{\partial}{\partial\phi_{2}}\,.
\end{eqnarray}
The corresponding negative-root generators are given by
\begin{equation}
E_{-\vec{\alpha}_{1}}=\left(E_{\vec{\alpha}_{1}}\right)^{*}\,,\quad
E_{-\vec{\alpha}_{2}}=-\left(E_{\vec{\alpha}_{2}}\right)^{*}\,,\quad
E_{-\vec{\alpha}_{3}}=-\left(E_{\vec{\alpha}_{3}}\right)^{*}\,.
\end{equation}
We shall not give the remaining generators. Note that the Cartan and
simple-root generators of the $SO(4)$ subgroup of $SO(2,4)$ are
$(H_2,H_3)$ and $(E_{\vec \alpha_2}, E_{\vec \alpha_3})$
respectively.

\bigskip
\noindent{\bf Scalar modes:}
\medskip

     In this case, we must have $s_1=s_2=0$, and the solution is given by
\begin{equation}
\Phi=e^{-{\rm i}E_{0}\tau}\left(\cosh\rho\right)^{-E_{0}}\,.
\end{equation}
The Laplacian action on $\Phi$ is given by
\begin{equation}
\Delta\Phi=-\Box\Phi=-\nabla_{\mu}\nabla^{\mu}\Phi=
-\frac{E_{0}\left(E_{0}-4\right)}{L^{2}}\Phi
\end{equation}
This scalar mode is a singlet under the $SO(4)$.

\bigskip
\noindent{\bf Vector modes:}
\medskip

For our choice of Cartan generators, we must have $s_1=1$ and
$s_2=0$.  The solution is given by
\begin{eqnarray}
A_{\tau} &=& e^{-{\rm i}\left(E_{0}\tau+\phi_{1}\right)}\sin\theta
\left(\cosh\rho\right)^{-E_{0}}\sinh\rho\,,\cr A_{\rho} &=& {\rm
i}\left(\cosh\rho\right)^{-1}\left(\sinh\rho\right)^{-1}A_{\tau}\,,\cr
A_{\theta} & = & {\rm
i}\left(\sin\theta\right)^{-1}\left(\cos\theta\right)A_{\tau}\,,\quad
A_{\phi_{1}}= A_{\tau}\,,\quad A_{\phi_{2}}  = 0\,.
\end{eqnarray}
The box and Laplacian operators acting on this solution are given by
\begin{equation}
\Box A_{\mu}=\frac{E_{0}^{2}-4E_{0}-1}{L^{2}}A_{\mu}\,,\qquad \Delta
A_{\mu}
=-\frac{\left(E_{0}-1\right)\left(E_{0}-3\right)}{L^{2}}A_{\mu}\,.
\end{equation}
Acting on this $(s_1,s_2)=(1,0)$ vector mode with
$E_{-\vec\alpha_2}$ and $E_{-\vec\alpha_3}$ repeatedly, we obtain
the four-dimensional vector representation of $SO(4)$, with the
weights given by $(\pm 1,0)$ and $(0,\pm 1)$, as shown in Fig.~1.

\bigskip
\noindent{\bf Spin-2 modes:}
\medskip

\begin{figure}[ht]
\subfloat[vector modes\label{fig:AdS5vector}]{\begin{centering}
\includegraphics[scale=0.4]{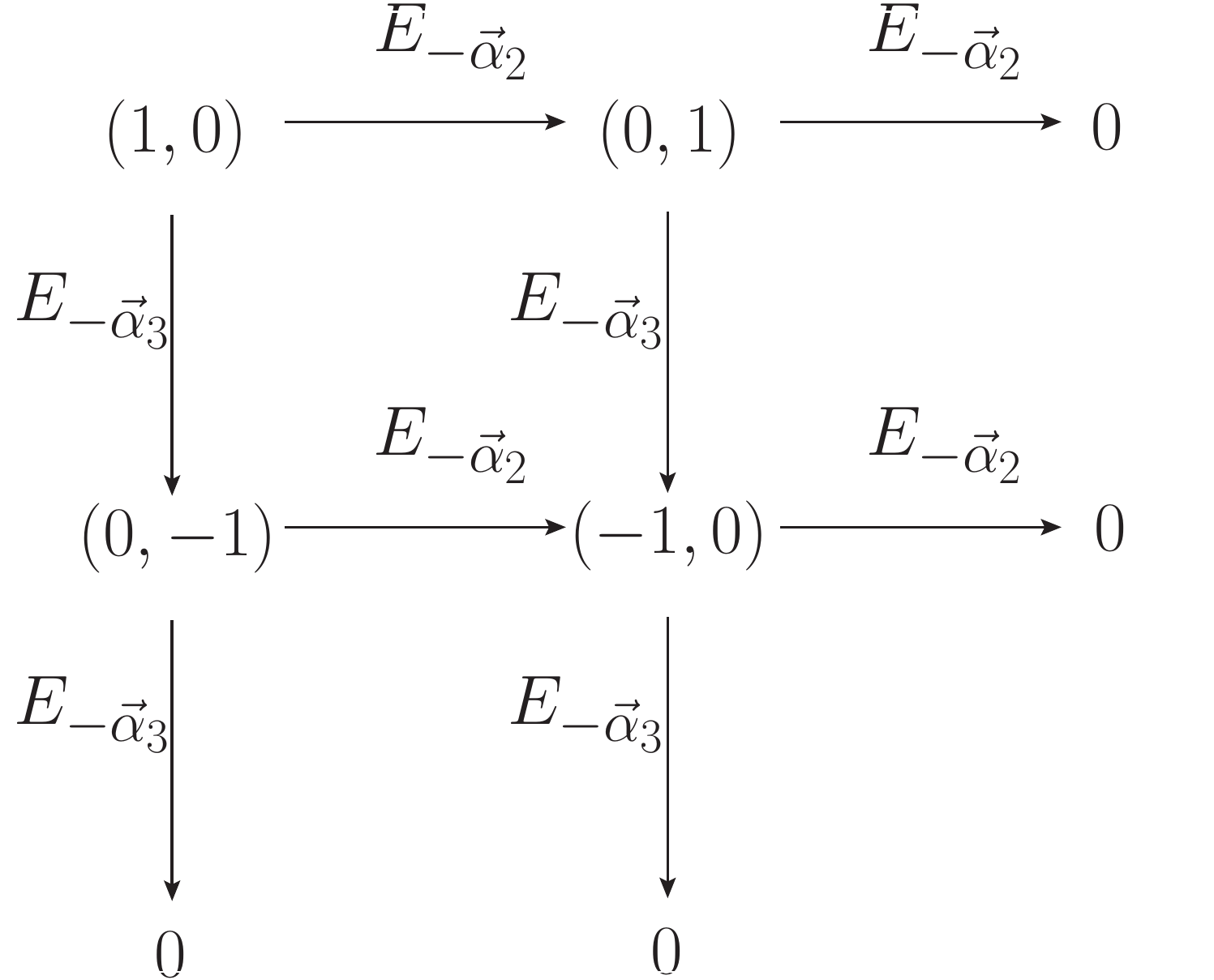}
\par\end{centering}}
\subfloat[Spin-2 modes\label{fig:AdS5Spin2}]{\begin{centering}
\includegraphics[scale=0.35]{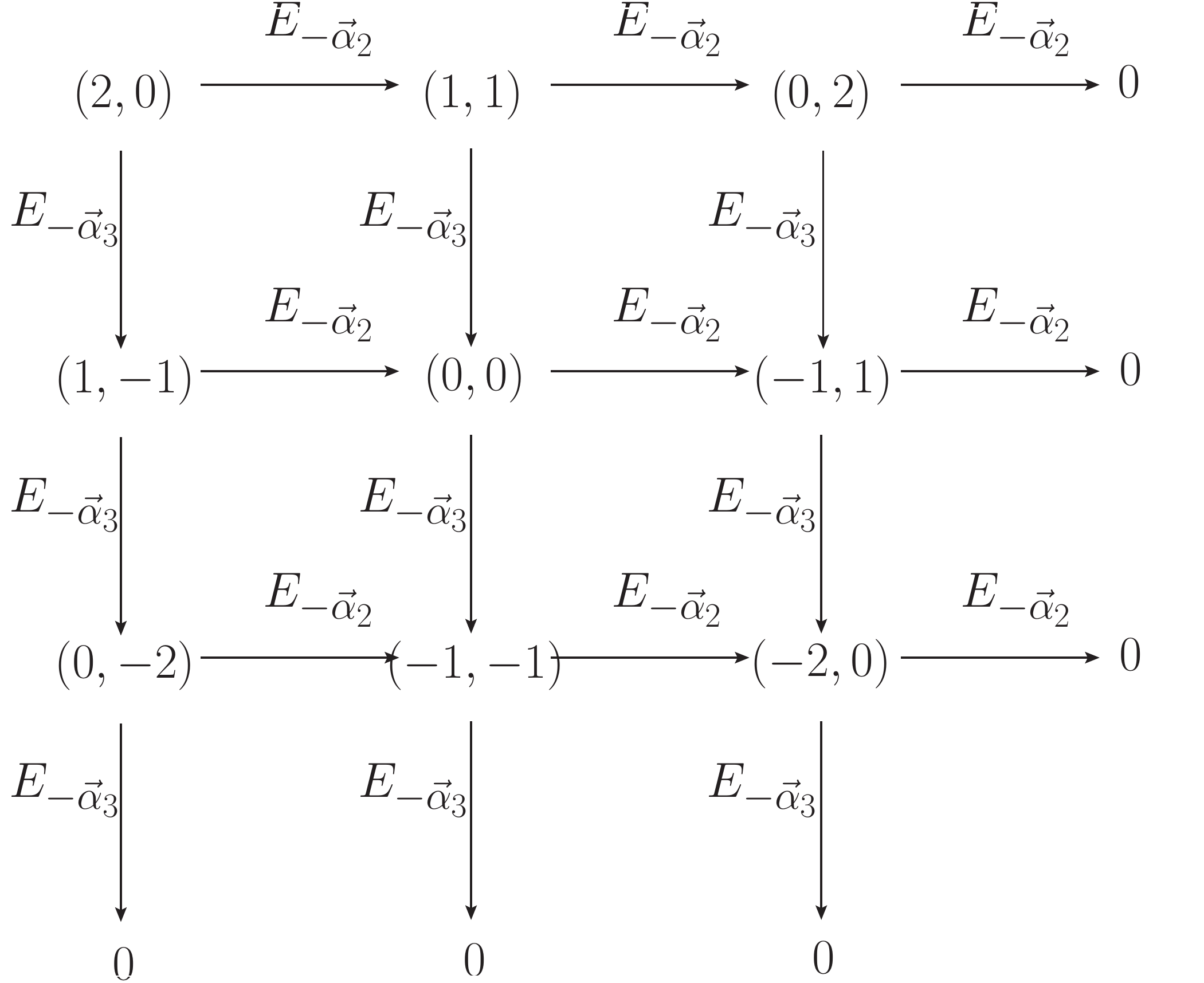}
\par\end{centering}}

\caption{AdS$_{5}$ modes: We present the weights of the full vector
and spin-2 representations of the $SO(4)$ subgroup of $SO(2,4)$, and
how these modes are constructed from their highest-weight states by
acting with the negative-root generators. \label{fig:AdS5modes}}
\end{figure}

     For this case, we must have $s_1=2$ and $s_2=0$.  The components of
the solution are given by
\begin{eqnarray}
\psi_{\tau\tau} & = & e^{-{\rm
i}\left(E_{0}\tau+2\phi_{1}\right)}\left(\sin\theta\right)^{2}
\left(\cosh\rho\right)^{-E_{0}}\left(\sinh\rho\right)^{2}\,,\cr
\psi_{\tau\rho} & = & {\rm i}\left(\cosh\rho\right)^{-1}
\left(\sinh\rho\right)^{-1}\psi_{\tau\tau} \,,\qquad
\psi_{\tau\theta} = {\rm i}\left(\sin\theta\right)^{-1}
\left(\cos\theta\right) \psi_{\tau\tau}\,,\cr \psi_{\tau\phi_{1}} &
= & \psi_{\tau\tau}\,,\qquad \psi_{\tau\phi_{2}} = 0\,,\qquad
\psi_{\rho\rho} = {\rm i}\left(\cosh\rho\right)^{-1}
\left(\sinh\rho\right)^{-1}\psi_{\tau\rho}\,,\cr \psi_{\rho\theta} &
= & {\rm i}\left(\sin\theta\right)^{-1}
\left(\cos\theta\right)\psi_{\tau\rho}\,,\qquad \psi_{\rho\phi_{1}}
= \psi_{\tau\rho}\,,\qquad \psi_{\rho\phi_{2}} = 0\,,\cr
\psi_{\theta\theta} & = & {\rm i}\left(\sin\theta\right)^{-1}
\left(\cos\theta\right)\psi_{\tau\theta}\,,\qquad
\psi_{\theta\phi_{1}} = \psi_{\tau\theta}\,,\qquad
\psi_{\theta\phi_{2}} = 0\,,\cr \psi_{\phi_{1}\phi_{1}} & = &
\psi_{\tau\phi_{1}}\,,\qquad \psi_{\phi_{1}\phi_{2}} = 0\,,\qquad
\psi_{\phi_{2}\phi_{2}}  = 0\,.
\end{eqnarray}
The box and Laplacian operators acting on this solution are given by
\begin{equation}
\Box\psi_{\mu\nu}=\frac{E_{0}^{2}-4E_{0}-2}{L^{2}}\psi_{\mu\nu}\,,\qquad
\Delta\psi_{\mu\nu} =-\frac{E_{0}^{2}-4E_{0}+8}{L^{2}}\psi_{\mu\nu}\,.
\end{equation}
Acting on this $(s_1,s_2)=(2,0)$ spin-2 mode with
$E_{-\vec\alpha_2}$ and $E_{-\vec\alpha_3}$ repeatedly, we obtain
the nine-dimensional spin-2 representation of $SO(4)$, with the
weights given by $(\pm 2,0)$, $(0,\pm 2)$, $(\pm 1, \pm 1)$ and
$(0,0)$.  The $\pm$ signs are independent. In
Fig.~\ref{fig:AdS5modes}, we present the construction of the vector
and spin-2 modes from their highest-weight states.

\subsection{AdS$_6$}

The AdS$_{6}$ metric is obtained by setting
$u_{1}=\sin\theta_{1}\sin\theta_{2}$,
$u_{2}=\sin\theta_{1}\cos\theta_{2}$, $u_{3}=\cos\theta_{1}$ in the
general metric (\ref{eq:AdSnmetric}).  It is given by
\begin{equation}
ds^{2}=L^{2}\left(-\cosh^{2}\rho
d\tau^{2}+d\rho^{2}+\sinh^{2}\rho\left(d\theta_{1}^{2}
+\sin^{2}\theta_{1}\left(d\theta_{2}^{2}
+\sin^{2}\theta_{2}d\phi_{1}^{2}+
\cos^{2}\theta_{2}d\phi_{2}^{2}\right)\right)\right)\,.
\end{equation}
The isometry group is $SO(2,5)$ and the three Cartan generators
are
\begin{equation}
H_{1}={\rm i}L_{12}={\rm i}\frac{\partial}{\partial\tau}\,,\quad
H_{2}={\rm i}L_{34}={\rm i}\frac{\partial}{\partial\phi_{1}}\,,\quad
H_{3}={\rm i} L_{56}={\rm i}\frac{\partial}{\partial\phi_{2}}\,.
\end{equation}
The simple roots are $\vec{\alpha}_{1}=\left(-1,-1,0\right)$,
$\vec{\alpha}_{2}=\left(0,1,-1\right)$, $\vec{\alpha}_{3}=
\left(0,0,1\right)$.  The corresponding generators are
\begin{eqnarray}
E_{\vec{\alpha}_{1}} & = & \ft{1}{2}\left(L_{13}+{\rm i} L_{14}+{\rm
i}L_{23}-L_{24}\right)\cr & = & \ft{1}{2}e^{{\rm
i}\left(\tau+\phi_{1}\right)}\sin\theta_{1}
\sin\theta_{2}\tanh\rho\frac{\partial}{\partial\tau} -\ft{1}{2}{\rm
i}e^{{\rm i}\left(\tau+\phi_{1}\right)}\sin\theta_{1}
\sin\theta_{2}\frac{\partial}{\partial\rho}\cr && -\ft{1}{2}{\rm
i}e^{{\rm i}\left(\tau+\phi_{1}\right)}
\cos\theta_{1}\sin\theta_{2}\coth\rho
\frac{\partial}{\partial\theta_{1}} -\ft{1}{2}{\rm i}e^{{\rm
i}\left(\tau+\phi_{1}\right)} \csc\theta_{1}\cos\theta_{2}\coth\rho
\frac{\partial}{\partial\theta_{2}}\cr && +\ft{1}{2}e^{{\rm
i}\left(\tau+\phi_{1}\right)} \csc\theta_{1}\csc\theta_{2}\coth\rho
\frac{\partial}{\partial\phi_{1}}\,,\cr E_{\vec{\alpha_{2}}} &=&
\ft{1}{2}\left(L_{35}+{\rm i}L_{36}-{\rm i}L_{45}+L_{46}\right)\cr
&=& -\ft{1}{2}e^{-{\rm i}\left(\phi_{1}-\phi_{2}\right)}
\frac{\partial}{\partial\theta_{2}} +\ft{1}{2}{\rm i}e^{-{\rm
i}\left(\phi_{1}-\phi_{2}\right)}
\cot\theta_{2}\frac{\partial}{\partial\phi_{1}} +\ft{1}{2}{\rm
i}e^{{\rm i}\left(\phi_{1}-\phi_{2}\right)}
\tan\theta_{2}\frac{\partial}{\partial\phi_{2}}\,,\cr
E_{\vec{\alpha}_{3}}&=& L_{57}-{\rm i}L_{67}\cr &=& -e^{-{\rm
i}\phi_{2}}\cos\theta_{2}\frac{\partial}{\partial\theta_{1}}
+e^{-{\rm i}\phi_{2}}\cot\theta_{1}\sin\theta_{2}
\frac{\partial}{\partial\theta_{2}}+{\rm i}e^{-{\rm i}\phi_{2}}
\cot\theta_{1}\sec\theta_{2}\frac{\partial}{\partial\phi_{2}}\,.
\end{eqnarray}
The corresponding negative-root generators are given by
\begin{equation}
E_{-\vec{\alpha}_{1}} =
\left(E_{\vec{\alpha}_{1}}\right)^{*}\,,\qquad
E_{-\vec{\alpha}_{2}}=-\left(E_{\vec{\alpha}_{2}}\right)^{*}\,,\qquad
E_{-\vec{\alpha}_{3}}=-\left(E_{\vec{\alpha}_{3}}\right)^{*}\,.
\end{equation}
The Cartan and the simple-root generators of the $SO(5)$ subgroup of
$SO(2,5)$ are $(H_2,H_3)$ and $(E_{\vec \alpha_2}, E_{\vec
\alpha_3})$ respectively.

\bigskip
\noindent{\bf Scalar modes:}
\medskip

    The non-trivial solution arises for $s_1=s_2=0$, and it is given by
\begin{equation}
\Phi=e^{-{\rm i}E_{0}\tau}\left(\cosh\rho\right)^{-E_{0}}\,.
\end{equation}
The Laplacian action on $\Phi$ is given by
\begin{equation}
\Delta\Phi=-\Box\Phi=-\nabla_{\mu}\nabla^{\mu}\Phi=
-\frac{E_{0}\left(E_{0}-5\right)}{L^{2}}\Phi
\end{equation}
As in the previous examples, this mode is a singlet under the
$SO(5)$.

\bigskip
\noindent{\bf Vector modes:}
\medskip

In this case, we must have $s_{1}=1$ and $s_{2}=0$, and the
corresponding solution is given by
\begin{eqnarray}
A_{\tau} & = & e^{-{\rm
i}\left(E_{0}\tau+\phi_{1}\right)}\sin\theta_{1}\sin\theta_{2}
\left(\cosh\rho\right)^{-E_{0}}\sinh\rho\,,\cr A_{\rho} & = & {\rm
i}\left(\cosh\rho\right)^{-1}\left(\sinh\rho\right)^{-1}A_{\tau}\,,\quad
A_{\theta_{1}} = {\rm
i}\left(\sin\theta_{1}\right)^{-1}\left(\cos\theta_{1}\right)A_{\tau}\,,\cr
A_{\theta_{2}} & = & {\rm
i}\left(\sin\theta_{2}\right)^{-1}\left(\cos\theta_{2}\right)A_{\tau}\,,\quad
A_{\phi_{1}}= A_{\tau}\,,\quad A_{\phi_{2}} = 0\,.
\end{eqnarray}
Note that for this solution, we have
\begin{equation}
\Box A_{\mu}=\frac{E_{0}^{2}-5E_{0}-1}{L^{2}}A_{\mu}\,,\qquad
\Delta A_{\mu}= -\frac{\left(E_{0}-1\right)\left(E_{0}-4\right)}{L^{2}}A_{\mu}\,.
\end{equation}
Acting with the negative-root generators of the $SO(5)$ subgroup, we
obtain the full five-dimensional spin-1 representation of $SO(5)$,
with the weights $(\pm 1, 0)$, $(0, \pm 1)$ and $(0,0)$, as shown in
Fig.~2.

\bigskip
\noindent{\bf Spin-2 modes:}
\medskip

\begin{figure}[ht]
\subfloat[vector
modes\label{fig:AdS6vector}]{\centering{}\includegraphics[scale=0.35]{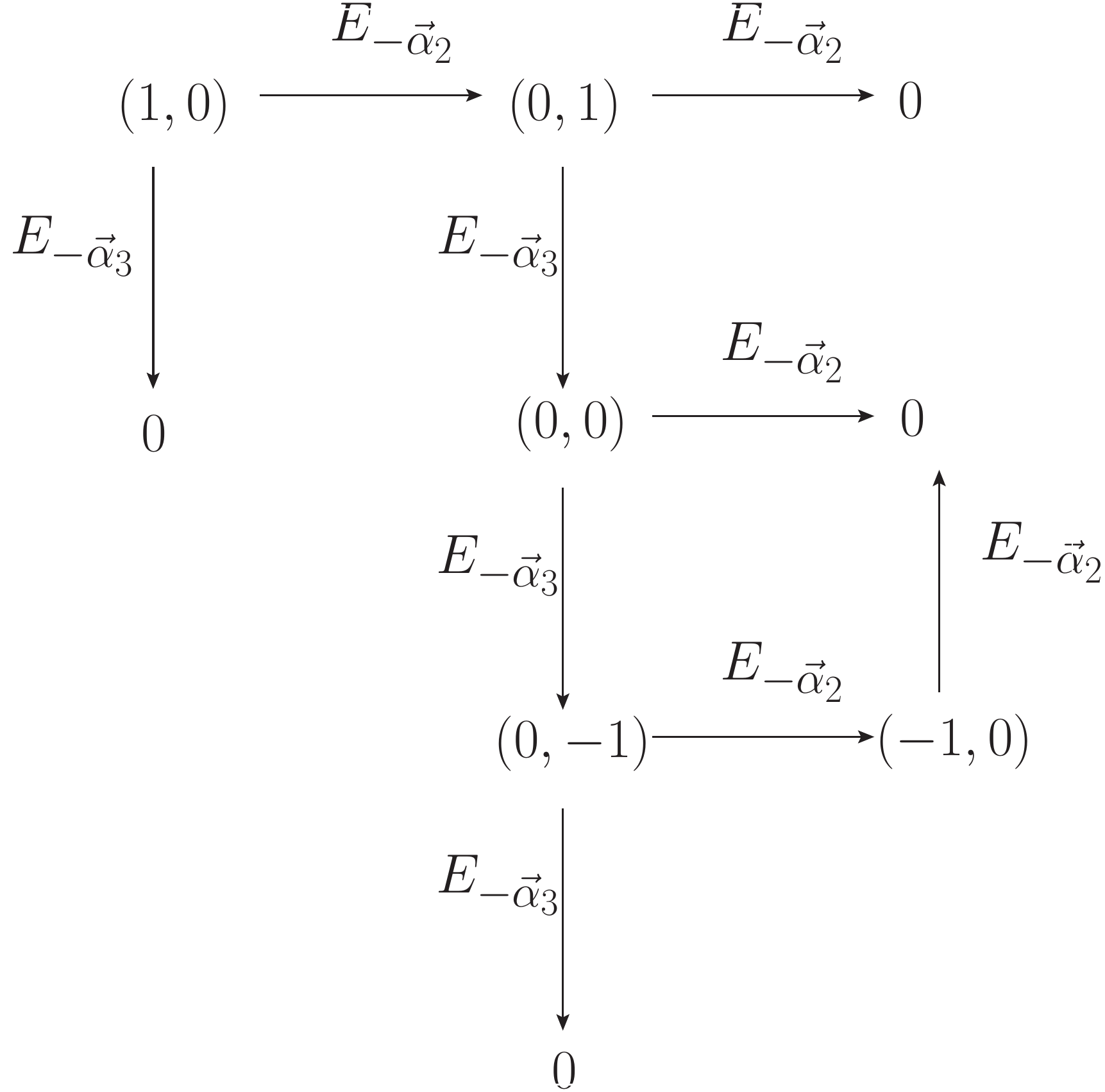}}
\subfloat[Spin-2
modes\label{fig:AdS6Spin2}]{\centering{}\includegraphics[scale=0.41]{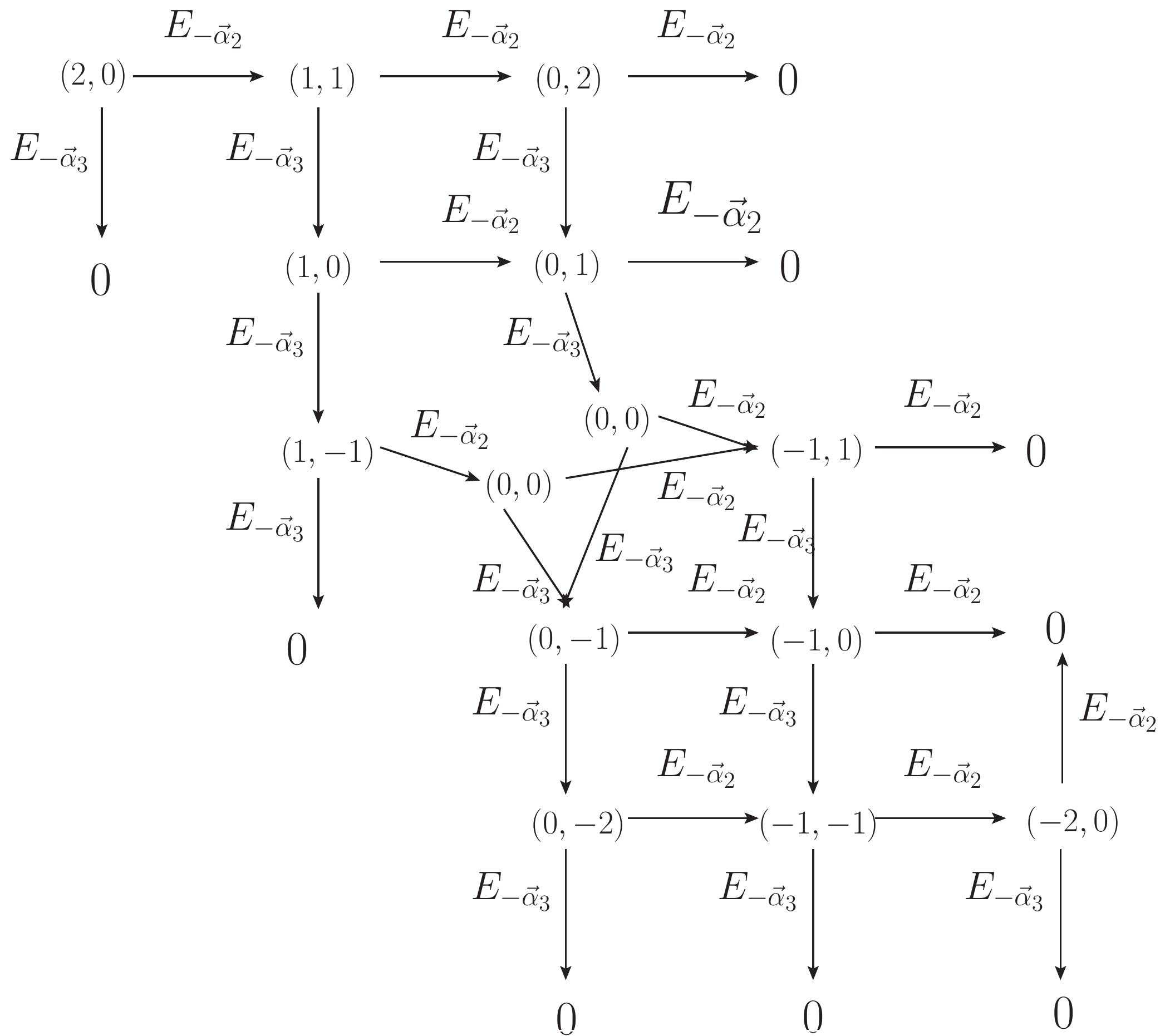}}

\caption{AdS$_{6}$ modes: We present the weights of the full vector
and spin-2 representations of the $SO(4)$ subgroup of $SO(2,4)$, and
how these modes are constructed from their highest weight states by
acting on the negative-root generators.\label{fig:AdS6modes}}
\end{figure}

    The non-trivial solution arises when $s_1=2$ and $s_2=0$, and it is
given by
\begin{eqnarray}
\psi_{\tau\tau} & = & e^{-{\rm
i}\left(E_{0}\tau+2\phi_{1}\right)}\left(\sin\theta_{1}\right)^{2}
\left(\sin\theta_{2}\right)^{2}\left(\cosh\rho\right)^{-E_{0}}
\left(\sinh\rho\right)^{2}\,,\cr \psi_{\tau\rho} & = & {\rm
i}\left(\cosh\rho\right)^{-1} \left(\sinh\rho\right)^{-1}
\psi_{\tau\tau}\,,\qquad \psi_{\tau\theta_{1}} = {\rm
i}\left(\sin\theta_{1}\right)^{-1}
\left(\cos\theta_{1}\right)\psi_{\tau\tau}\,,\cr
\psi_{\tau\theta_{2}} &=& {\rm i}\left(\sin\theta_{2}\right)^{-1}
\left(\cos\theta_{2}\right)\psi_{\tau\tau}\,,\qquad
\psi_{\tau\phi_{1}} = \psi_{\tau\tau}\,,\qquad \psi_{\tau\phi_{2}} =
0\,,\cr
\psi_{\rho\rho} & = & {\rm i}\left(\cosh\rho\right)^{-1}
\left(\sinh\rho\right)^{-1}\psi_{\tau\rho}\,,\qquad
\psi_{\rho\theta_{1}} = i\left(\sin\theta_{1}\right)^{-1}
\left(\cos\theta_{1}\right)\psi_{\tau\rho}\,,\cr
\psi_{\rho\theta_{2}} & = & {\rm i}\left(\sin\theta_{2}\right)^{-1}
\left(\cos\theta_{2}\right)\psi_{\tau\rho}\,,\qquad
\psi_{\rho\phi_{1}}=\psi_{\tau\rho}\,,\qquad
\psi_{\rho\phi_{2}}=0\,,\cr
\psi_{\theta_{1}\theta_{1}} & = & {\rm i}
\left(\sin\theta_{1}\right)^{-1}
\left(\cos\theta_{1}\right)\psi_{\tau\theta_{1}}\,,\qquad
\psi_{\theta_{1}\theta_{2}} =i\left(\sin\theta_{2}\right)^{-1}
\left(\cos\theta_{2}\right)\psi_{\tau\theta_{1}}\,,\cr 
\psi_{\theta_{1}\phi_{1}} & = & \psi_{\tau\theta_{1}}\,,\qquad
\psi_{\theta_{1}\phi_{2}} = 0\,,\qquad \psi_{\theta_{2}\theta_{2}} =
{\rm i}\left(\sin\theta_{2}\right)^{-1}
\left(\cos\theta_{2}\right)\psi_{\tau\theta_{2}}\,,\cr
\psi_{\theta_{2}\phi_{1}} & = & \psi_{\tau\theta_{2}}\,,\qquad
\psi_{\theta_{2}\phi_{2}} = 0\,,\qquad
\psi_{\phi_{1}\phi_{1}} =\psi_{\tau\phi_{1}}\,,\cr
\psi_{\phi_{1}\phi_{2}} & = & 0\,,\qquad
\psi_{\phi_{2}\phi_{2}}= 0\,.
\end{eqnarray}
The solution satisfies
\begin{equation}
\Box\psi_{\mu\nu}=\frac{E_{0}^{2}-5E_{0}-2}{L^{2}}\psi_{\mu\nu}\,,\qquad
\Delta\psi_{\mu\nu}= -\frac{E_{0}^{2}-5E_{0}+10}{L^{2}}\psi_{\mu\nu}\,.
\end{equation}
Acting with the negative-root generators of the $SO(5)$ subgroup, we
obtain the full 14-dimensional spin-2 representation of $SO(5)$,
with the weights $(\pm 2, 0)$, $(0, \pm 2)$, $(\pm 1,\pm 1)$, $(\pm
1, 0)$, $(0,\pm 1)$ and $(0,0)_2$, where the subscript denotes the
degeneracy and $\pm$ are independent. In Fig.~\ref{fig:AdS6modes},
we present the construction of the vector and spin-2 modes from
their highest-weight states.

\section{Linearized modes in general AdS$_D$}

      In the previous sections, we obtain explicit solutions for the
linearized scalar, vector and spin-2 modes in AdS spactimes in
$D=3,4,5$ and 6.  The results can be easily generalized to arbitrary
dimensions.  In this section, we present the general solutions,
which we verify up to $D=11$.

\subsection{$D=2n+1$}

The metric is given by (\ref{eq:AdSnmetric}) with $i=1,2,\ldots,n$.
We parameterize $u_i$ as follows
\begin{eqnarray}
u_{1} & = & \sin\theta_{1}\sin\theta_{2}\cdots\sin\theta_{n-1}\,,\cr
u_{2} & = & \sin\theta_{1}\sin\theta_{2}\cdots\cos\theta_{n-1}\,,\cr
u_{3} & = & \sin\theta_{1}\sin\theta_{2}\cdots\cos\theta_{n-2}\,,\cr
 & \cdots\cr
u_{n} & = & \cos\theta_{1}\,.\label{oddDui}
\end{eqnarray}
To be explicit, we have
\begin{eqnarray}
ds^{2} &=& L^{2}\Biggl(-\cosh^{2}\rho
d\tau^{2}+d\rho^{2}+\sinh^{2}\rho\biggl(d\theta_{1}^{2}
+\cos\theta_{1}d\phi_{n}^{2}+\sin^{2}\theta_{1}
\Bigl(d\theta_{2}^{2}+\cos^{2}\theta_{2}d\phi_{n-1}^{2}\cr &&
+\sin^{2}\theta_{2}\bigl(\cdots(d\theta_{n-1}^{2}+
\cos\theta_{n-1}d\phi_{2}^{2}+\sin\theta_{n-1}
d\phi_{1}^{2})\bigr)\Bigr)\biggr)\Biggr)\,.
\end{eqnarray}
We shall not present the $SO(2,2n)$ generators, discussed in section
2, in these coordinates, but give the solutions directly.  As in the
previous sections, we give only the solutions that are the
highest-weight states of the $SO(2,2n)$.

\bigskip
\noindent{\bf Scalar modes:}
\medskip

The solution requires that $s_i=0$, and it is given by
\begin{equation}
\Phi=e^{-{\rm
i}E_{0}\tau}\left(\cosh\rho\right)^{-E_{0}}\,.\label{oddDscalar}
\end{equation}
The Laplacian acting on $\Phi$ is
\begin{equation}
\Delta\Phi=-\Box\Phi=-\frac{E_{0}\left(E_{0}-D+1\right)}{L^{2}}\Phi\,.
\label{oddDscalar1}
\end{equation}

\bigskip
\noindent{\bf Vector modes:}
\medskip

    In this case, we must have $s_1=1$ and $s_i=0$ for $i\ge2$. The
solution is given by
\begin{eqnarray}
A_{\tau} & = & e^{-{\rm
i}\left(E_{0}\tau+\phi_{1}\right)}\sin\theta_{1}
\sin\theta_{2}\cdots\sin\theta_{n-1}
\left(\cosh\rho\right)^{-E_{0}}\sinh\rho\,,\cr A_{\rho} & = & {\rm
i}\left(\cosh\rho\right)^{-1} \left(\sinh\rho\right)^{-1}
A_{\tau}\,,\qquad A_{\theta_{i}} = {\rm
i}\left(\sin\theta_{i}\right)^{-1} \left(\cos\theta_{i}\right)
A_{\tau}\,,\cr 
A_{\phi_{1}} & = & A_{\tau}\,,\qquad A_{\phi_{2}} =
\cdots=A_{\phi_{n}}=0\,.\label{oddDvector}
\end{eqnarray}
It satisfies
\begin{equation}
\Box A_{\mu}=\frac{E_{0}^{2}-(D-1)E_{0}-1}{L^{2}}A_{\mu}\,,\qquad
\Delta A_{\mu}= -\frac{\left(E_{0}-1\right)
\left(E_{0}-\left(D-2\right)\right)}{L^{2}}A_{\mu}\,.\label{oddDvector1}
\end{equation}

\bigskip
\noindent{\bf Spin-2 modes:}
\medskip

In this case, we have $s_1=2$ and $s_i=0$ for $i\ge 2$.  The solution is given by
\begin{eqnarray}
\psi_{\tau\tau} & = & e^{-{\rm i}\left(E_{0}\tau+2\phi_{1}\right)}
\left(\sin\theta_{1} \sin\theta_{2}\cdots
\sin\theta_{n-1}\right)^{2} \left(\cosh\rho\right)^{-E_{0}}
\left(\sinh\rho\right)^{2}\,,\cr \psi_{\tau\rho} & = & {\rm
i}\left(\cosh\rho\right)^{-1}
\left(\sinh\rho\right)^{-1}\psi_{\tau\tau}\,,\qquad
\psi_{\tau\theta_{i}} = {\rm i}\left(\sin\theta_{i}\right)^{-1}
\left(\cos\theta_{i}\right)\psi_{\tau\tau}\,,\cr 
\psi_{\tau\phi_{1}} & = & \psi_{\tau\tau}\,,\qquad
\psi_{\tau\phi_{2}}  = \cdots=\psi_{\tau\phi_{n}}=0\,,\cr 
\psi_{\rho\rho} &=& {\rm i}\left(\cosh\rho\right)^{-1}
\left(\sinh\rho\right)^{-1} \psi_{\tau\rho} \,, \qquad
\psi_{\rho\theta_{i}} = {\rm i}\left(\sin\theta_{i}\right)^{-1}
\left(\cos\theta_{i}\right)\psi_{\tau\rho}\,,\cr 
\psi_{\rho\phi_{1}} & = & \psi_{\tau\rho}\,,\qquad
\psi_{\rho\phi_{2}}  = \cdots=\psi_{\rho\phi_{n}}=0\,,\cr 
\psi_{\theta_{i}\theta_{i}} & = & {\rm
i}\left(\sin\theta_{i}\right)^{-1}
\left(\cos\theta_{i}\right)\psi_{\tau\theta_{i}}\,,\qquad
\psi_{\theta_{i}\theta_{j}} = {\rm i}\left(\sin\theta_{j}
\right)^{-1} \left(\cos\theta_{j}\right)\psi_{\tau\theta_{i}}
\,,\cr
\psi_{\theta_{i}\phi_{1}} & = & \psi_{\tau\theta_{i}}\,,\quad
\psi_{\theta_{i}\phi_{2}} =\cdots=\psi_{\theta_{i}\phi_{n}}=0\,,
\quad \psi_{\phi_{1}\phi_{1}} =\psi_{\tau\phi_{1}}\,,\cr 
\psi_{\phi_{1}\phi_{2}} & = &
\cdots=\psi_{\phi_{1}\phi_{n}}=\psi_{\phi_{2}\phi_{2}}=
\cdots=\psi_{\phi_{2}\phi_{n}}=\cdots=\psi_{\phi_{n}\phi_{n}}=0\,.
\label{oddDspin2}
\end{eqnarray}
The box and Laplacian action on this solution is given by
\begin{equation}
\Box\psi_{\mu\nu}=\frac{E_{0}^{2}-(D-1)E_{0}-2}{L^{2}}\psi_{\mu\nu}\,,\qquad
\Delta\psi_{\mu\nu} =
-\frac{E_{0}^{2}-(D-1)E_{0}+2(D-1)}{L^{2}}\psi_{\mu\nu}\,.\label{oddDspin21}
\end{equation}

\subsection{$D=2n$}

The AdS$_{2n}$ metric is given by (\ref{eq:AdSnmetric}) with $u_i$
also given by (\ref{oddDui}). To be explicit, we have
\begin{eqnarray}
ds^{2} & = & L^{2}\Biggl(-\cosh^{2}\rho
d\tau^{2}+d\rho^{2}+\sinh^{2}\rho
\biggl(d\theta_{1}^{2}+\sin^{2}\theta_{1}
\Bigl(d\theta_{2}^{2}+\cos^{2}\theta_{2}d\phi_{n-1}^{2}\cr &&
+\sin^{2}\theta_{2}\bigl(\cdots(d\theta_{n-1}^{2}
+\cos\theta_{n-1}d\phi_{2}^{2}+\sin\theta_{n-1}d\phi_{1}^{2})
\bigr)\Bigr)\biggr)\Biggr)\,.
\end{eqnarray}
Comparing to the AdS$_{2n+1}$ metric, there are no $\phi_{n}$
coordinate in AdS$_{2n}$. Thus the scalar modes take the identical
form as (\ref{oddDscalar}).  The vector and spin-2 modes can also be
read off from those in odd dimensions (\ref{oddDvector}) and
(\ref{oddDspin2}) respectively, by ignoring the $\phi_n$ components,
which vanish anyway.  The box and Laplacian action on these modes
are given by (\ref{oddDscalar1}), (\ref{oddDvector1}) and
(\ref{oddDspin21}), but with $D=2n$.

Having obtained the highest weight states, we can obtain the
remaining modes in the representation of the subgroup $SO(D-1)$ of
$SO(2,D-1)$.  The scalar mode is a singlet under the $SO(D-1)$.  The
vector modes form $(D-1)$-dimensional spin-1 representation. For
$D=2n+1$, the weights are $n$-vectors, $(0,\ldots, \pm 1, \ldots,
0)$.  For $D=2n+2$, an extra weight vector arises, namely
$(0,\ldots,0)$.  The spin-2 modes, on the other hand, form the
spin-2 representation of $SO(D-1)$ with dimensions $\fft12
D(D-1)-1$.  For $D=2n+1$, the weights are $(0,\ldots, \pm 2, \ldots,
0)$, $(0,\ldots, \pm 1, \ldots,\pm 1,\ldots,0)$ and
$(0,\ldots,0)_{n-1}$, where $\pm$ are independent and the subscript
denotes the degeneracy.  In $D=2n+2$, additional weights,
$(0,\ldots, \pm 1, \ldots, 0)$ and $(0,\ldots,0)$ arise.

\section{Log modes}

In higher-derivative extended gravities, for appropriate choice of
parameters, the differential operator may factorize into the follow
form
\begin{equation}
(\widetilde \Box - m_1^2)\cdots (\widetilde \Box -m_n^2)\psi=0\,.
\end{equation}
where $\widetilde\Box\equiv \Box + c\, \Lambda$ for appropriate
constant $c$ and $\widetilde\Box \psi_0=0$ define a massless mode
$\psi_0$.  (See, for example, \cite{lpcritical,lpp}.)  Thus the most
general solution is the linear combination of all the modes
associated with mass $m_n$.  A critical point is defined when two
mass parameters become coincident, in which case log modes emerge.
The log modes are defined by
\begin{equation}
(\widetilde \Box - m_1^2)^2 \psi_{\rm log}=0\,,\qquad
\hbox{but}\qquad (\widetilde\Box - m_1^2)_{\rm log}\ne 0\,.
\end{equation}
From the general massive solutions we obtained in the previous
section, we can easily obtain the log modes, namely
\begin{equation}
\psi_{\rm log} = \fft{\partial \psi(m^2)}{\partial
(m^2)}\Big|_{m\rightarrow m_1}\,.
\end{equation}
Since all the solutions have a universal factor $e^{-{\rm i} E_0
\tau} (\cosh\rho)^{-E_0}$, it is easy to see that the log modes are
proportional to the corresponding massive or massless modes with an
overall factor
\begin{equation}
f=-{\rm i}\, \tau - \log (\cosh\rho)\,.
\end{equation}
The log modes were obtained in \cite{Grumiller:2008qz} in three
dimensions and \cite{bhrt} in four dimensions. When there are
multiple $m_i$ that are coincident, so that the equation becomes
\begin{equation}
(\widetilde \Box - m_1)^k\psi =0\,.\label{kd}
\end{equation}
The most general solution of (\ref{kd}) is then given by
\begin{equation}
\psi = \sum_{i=0}^{k-1} c_i \fft{\partial^{i}\psi(m^2)}{\partial
(m^2)^i}\Big|_{m\rightarrow m_1}=\Big(\sum_{i=0}^{k-1} c_i f^i\Big)
\psi(m_1^2) \,.
\end{equation}

\section{Conclusions and discussions}

In this paper, we have constructed the general massless and massive
scalar, vector and spin-2 modes in AdS vacua in diverse dimensions.
These modes may arise in extended and critical gravities with
higher-order curvature terms.  As we have mentioned in the
introduction, the explicit solutions may yield useful information of
the relevant theories. An important property of these modes is the
following constraint
\begin{equation}
M_s^2 = \fft{E_0^2 - (D-1) E_0}{L^2}\,,\label{MsE0}
\end{equation}
where $M_s^2$, defined in (\ref{spinseom}), denotes $M_0^2$, $M_1^2$
and $M_2^2$ that appear in (\ref{spin012eom}).  The reality
condition for $E_0$, which ensures that there is no expoential
growth in time, is given by
\begin{equation}
M_s^2 \ge M_{\rm BF}^2 \equiv -\fft{(D-1)^2}{4L^2}\,.
\end{equation}
Thus, we see that there is a universal Breitenlohner-Freedman bound
for the scalar, vector and the spin-2 fields.  We expect that this
result generalizes to higher spin-$s$ fields.  Note that in the case
of topologically massive gravity, the operator is factorized to
three linear factors with no square root parameters, and hence there
is no such a bound.

As we have explained in section 2, the parameter $M_s$ does not
necessarily denote the true mass of the modes.  In order to find out
the proper definition of the mass, we need to define the meaning of
masslessness in the AdS spacetimes.  Let $r=\sinh\rho$ denote the
radius of the foliating sphere in the AdS given in
(\ref{eq:AdSnmetric}), we find that for large $r$, the spin-$s$
modes we obtained fall off as follows
\begin{equation}
\psi_s \sim \fft{1}{r^{E_0-s}}\,.
\end{equation}
It is natural to expect that all the massless modes should fall off
the same way as the Schwarzschild-AdS black holes.  Thus for
massless modes we have
\begin{equation}
E_0 = D-3 + s\,.\label{masslessE0}
\end{equation}
It follows that a massless spin-$s$ mode $\psi_s^{(0)}$ satisfies
\begin{equation}
\Big(\Box + \fft{2 - (s-2)(D+s-4)}{L^2}\Big)\psi_s^{(0)}=0\,.
\end{equation}
The corresponding parameter $M_s^2$ is given by
\begin{equation}
M_s^2 = (M_s^{(0)})^2\equiv \fft{(s-2)(D+s-3)}{L^2} = M_{\rm BF}^2 +
\fft{(D+2s-5)^2}{L^2} \ge M_{\rm BF}^2\,.\label{masslessmass}
\end{equation}
We see that the massless modes always satisfy the tachyon-free
condition. Note that this definition of massless fields coincides
precisely the one discussed in section 2 for $s=1,2$.  For the
scalar $s=0$ field, it follows from (\ref{masslessmass}) that the
massless mode is given by $(M_0^{(0)})^2= -2(D-3)$. We verify, with
many examples of supergravities ({\it e.g.}, \cite{10author},) that
this definition of massless scalar is consistent with supergravity
scalar potentials.  (Scalars that belong to the graviton
super-multiplet in supergravities are naturally massless.) It
follows that the true mass of the spin-$s$ solutions we obtained is
given by
\begin{equation}
{\cal M}_s^2 = M_s^2 - (M_s^{(0)})^2 = \fft{(E_0 - D+3
-s)(E_0-2+s)}{L^2}\,.
\end{equation}
The true mass ${\cal M}_s$ of a spin-$s$ field in the AdS background
can thus be defined by
\begin{equation}
\Big(\Box + \fft{2 - (s-2)(D+s-4)}{L^2} - {\cal
M}_s^2\Big)\psi_s=0\,.
\end{equation}
In terms of this properly defined mass ${\cal M}_s$, the generalized
Breitenlohner-Freedman bound is now given by
\begin{equation}
({\cal M}_s^{\rm BF})^2 \ge - \fft{(D+2s-5)^2}{4L^2}\,.
\end{equation}

    Having obtained the generalized Breitenlohner-Freedman bound that
ensures tachyon free, we examine the falloffs of the modes at the
AdS boundary.  It follows from (\ref{MsE0}) and (\ref{masslessE0})
that if $D-5 +2s\le 0$, the modes $\psi_s$ have falloffs no slower
than the Schwarzschild black hole at the AdS boundary.  For modes
with $D-5+2s > 0$, which include the spin-2 modes, two situations
can occur.  The first-type of modes have $M_s^2\ge (M_s^{(0)})^2$,
corresponding to $E_0 \ge D-3 + s$, they satisfy the standard AdS
boundary condition. (The $E_0 \le 2-s$ branch violates the boundary
condition.)  The second-type of modes have $M^2_s$ lie in the range,
\begin{equation}
M_{\rm BF}^2 \le M_s^2 < (M_s^{(0)})^2\,.
\end{equation}
They have slower falloffs than the standard AdS boundary condition.
These modes should be truncated out from the spectrum.  Our explicit
construction of the linearized modes confirms the conclusion of
\cite{lpp}.

     To conclude, the explicit construction of the linearized modes
that could arise in extended and critical gravities with
higher-derivative curvature terms should be served as useful tool to
study various properties of these theories.

\section*{Acknowledgement}

   We are grateful to Haishan Liu, Yi Pang and Chris Pope for useful
discussions.  grant acknowledgement also.  The research of
Y.X.C.~and K.N.S.~is supported in part by NSFC grant 11075138, and
973-Program grant 2005CB724508.  The research of H.L.~is supported
in part by the NSFC grant 11175269.


\begin{thebibliography}{99}
\bibitem{stelle1} K.S. Stelle,
{\it Renormalization of higher derivative quantum gravity}, Phys.
Rev. {\bf D16}, 953 (1977).

\bibitem{stelle2} K.S. Stelle,
{\it Classical gravity with higher derivatives}, Gen. Rel. Grav.
{\bf 9}, 353 (1978).

\bibitem{tmg} S. Deser, R. Jackiw and S. Templeton,
{\it Topologically massive gauge theories}, Annals Phys. {\bf 140},
372 (1982);
S. Deser, {\it Cosmological topological supergravity}, in
Christensen, S.M. (Ed.): Quantum Theory Of Gravity, 374-381 (1982).

\bibitem{nmg} E.A. Bergshoeff, O. Hohm and P.K. Townsend,
{\it Massive gravity in three dimensions,}
  Phys. Rev. Lett.  {\bf 102}, 201301 (2009), arXiv:0901.1766 [hep-th].

\bibitem{lss} W. Li, W. Song and A. Strominger,
{\it Chiral gravity in three dimensions}, JHEP {\bf 0804}, 082
(2008), arXiv:0801.4566 [hep-th].

\bibitem{lpcritical}
  H. L\"u and C.N. Pope,
{\it Critical gravity in four dimensions,} Phys. Rev. Lett.  {\bf
106}, 181302 (2011), arXiv:1101.1971 [hep-th].

\bibitem{maldaconf}
  J. Maldacena,
{\it Einstein gravity from conformal gravity,} arXiv:1105.5632
[hep-th].

\bibitem{lpp}
  H.~Lu, Y.~Pang and C.N.~Pope,
{\it Conformal gravity and extensions of critical gravity,}
  arXiv:1106.4657 [hep-th], to appear in Phys. Rev. D.

\bibitem{lpsw}
  H.~L\"u, C.~N.~Pope, E.~Sezgin and L.~Wulff,
{\it Critical and non-critical Einstein-Weyl supergravity,}
  arXiv:1107.2480 [hep-th].

\bibitem{Andringa:2009yc}
  R.~Andringa, E.A.~Bergshoeff, M.~de Roo, O.~Hohm, E.~Sezgin and P.K.~Townsend,
{\it Massive $3D$ supergravity,}
  Class.\ Quant.\ Grav.\  {\bf 27}, 025010 (2010)
  [arXiv:0907.4658 [hep-th]].

\bibitem{bhrst}
  E.A.~Bergshoeff, O.~Hohm, J.~Rosseel, E.~Sezgin and P.K.~Townsend,
{\it More on massive $3D$ supergravity,}
  Class.\ Quant.\ Grav.\  {\bf 28}, 015002 (2011)
  [arXiv:1005.3952 [hep-th]].

\bibitem{Bergshoeff:2010ui}
  E.~A.~Bergshoeff, O.~Hohm, J.~Rosseel and P.~K.~Townsend,
{\it On maximal massive $3D$ supergravity,}
  Class.\ Quant.\ Grav.\  {\bf 27}, 235012 (2010)
  [arXiv:1007.4075 [hep-th]].

\bibitem{Lu:2010ct}
  H.~L\"u, C.N.~Pope and E.~Sezgin,
{\it Massive three-dimensional supergravity from $R + R^2$ action in
six dimensions,}
  JHEP {\bf 1010}, 016 (2010)
  [arXiv:1007.0173 [hep-th]].

\bibitem{Lu:2010cg}
  H.~L\"u, Y.~Pang,
{\it On hybrid (topologically) massive supergravity in three
dimensions,} JHEP {\bf 1103}, 050 (2011).
  [arXiv:1011.6212 [hep-th]].

\bibitem{liusun} Y. Liu and Y.W. Sun,
{\it Note on new massive gravity in AdS$_3$,}
  JHEP {\bf 0904}, 106 (2009), arXiv:0903.0536 [hep-th].

\bibitem{dllpst}
  S.~Deser, H.~Liu, H.~L\"u, C.N.~Pope, T.C.~Sisman and B.~Tekin,
{\it Critical points of $D$-dimensional extended gravities,}
  Phys.\ Rev.\  D {\bf 83}, 061502 (2011)
  [arXiv:1101.4009 [hep-th]].

\bibitem{Pang}
  Y.~Pang,
{\it Brief note on AMD conserved quantities in quadratic curvature
theories,}
  Phys.\ Rev.\  D {\bf 83}, 087501 (2011)
  [arXiv:1101.4267 [hep-th]].

\bibitem{Alishahiha:2011yb}
  M.~Alishahiha and R.~Fareghbal,
{\it $D$-dimensional log gravity,}
  Phys.\ Rev.\  D {\bf 83}, 084052 (2011)
  [arXiv:1101.5891 [hep-th]].

\bibitem{Gullu:2011sj}
  I.~Gullu, M.~Gurses, T.C.~Sisman and B.~Tekin,
{\it AdS waves as exact solutions to quadratic gravity,}
  Phys.\ Rev.\  D {\bf 83}, 084015 (2011)
  [arXiv:1102.1921 [hep-th]].

\bibitem{bhrt}
  E.A.~Bergshoeff, O.~Hohm, J.~Rosseel and P.K.~Townsend,
{\it Modes of log gravity,}
  Phys.\ Rev.\  D {\bf 83}, 104038 (2011)
  [arXiv:1102.4091 [hep-th]].

\bibitem{AyonBeato:2011qw}
  E.~Ayon-Beato, G.~Giribet and M.~Hassaine,
{\it Deeper discussion of Schr\"odinger invariant and logarithmic
sectors of higher-curvature gravity,}
  Phys.\ Rev.\  D {\bf 83}, 104033 (2011)
  [arXiv:1103.0742 [hep-th]].

\bibitem{Porrati:2011ku}
  M.~Porrati and M.~M.~Roberts,
{\it Ghosts of critical gravity,}
  Phys.\ Rev.\  D {\bf 84}, 024013 (2011)
  [arXiv:1104.0674 [hep-th]].

\bibitem{Liu:2011kf}
  H.~Liu, H.~L\"u and M.~Luo,
{\it On black hole stability in critical gravities,}
  arXiv:1104.2623 [hep-th].

\bibitem{Myung:2011uy}
  Y.S.~Myung, Y.W.~Kim and Y.J.~Park,
{\it Critical gravity on AdS$_2$ spacetimes,}
  arXiv: 1106.0546 [hep-th].

\bibitem{Bergshoeff:2011xy}
  E.A.~Bergshoeff, S.~de Haan, W.~Merbis and J.~Rosseel,
{\it A non-relativistic logarithmic conformal field theory from a
holographic point of view,}
  arXiv:1106.6277 [hep-th].

\bibitem{Myung:2011nn}
  Y.S.~Myung,
{\it Critical gravity as van Dam-Veltman-Zakharov discontinuity in
anti de Sitter space,}
  arXiv:1107.3594 [hep-th].

\bibitem{Moon:2011ef}
  T.~Moon and Y.S.~Myung,
{\it Critical gravity in the Chern-Simons modified gravity,}
  arXiv:1108. 2612 [hep-th].

\bibitem{10author}
M.~Cveti\v c, M.J. Duff, P. Hoxha, J.T. Liu, H. L\"u, J.X. Lu, R.
Martinez-Acosta, C.N. Pope, H. Sati, T.A. Tran, {\it Embedding AdS
black holes in ten-dimensions and eleven-dimensions,}
  Nucl.\ Phys.\  B {\bf 558}, 96 (1999)
  [arXiv:hep-th/9903214].

\bibitem{Grumiller:2008qz}
  D.~Grumiller and N.~Johansson,
  {\it Instability in cosmological topologically massive gravity at the chiral
  point},
  JHEP {\bf 0807}, 134 (2008)
  [arXiv:0805.2610 [hep-th]].

\end{thebibliography}
\end{document}